\documentclass[aps,prd,superscriptaddress,nofootinbib,amsmath,amssymb]{revtex4}
\usepackage{graphicx}
\usepackage{dcolumn}
\usepackage{bm,latexsym}
\usepackage{mathrsfs}

\def\Journal#1#2#3#4{{#1} {\bf #2}, #3 (#4)}

\def\APJ{{Astrophys. J.}}

\def\CQG{{Class. Quant. Grav.}}

\def\JMP{{J. Math. Phys.}}
\def\LivRev{{Living Rev. Relativity}} 
\def\PLB{{Phys. Lett.}  B}

\def\PRL{Phys. Rev. Lett.}
\def\PREV{Phys. Rev.}
\def\PRD{{Phys. Rev.} D}

\def\RMF{{Rev. Mex. F\'\i s.}}

\def\be{\begin{equation}}
\def\ee{\end{equation}}
\def\bea{\begin{eqnarray}}
\def\eea{\end{eqnarray}}

\begin{document}

\title{The initial value problem of scalar-tensor theories of gravity}
\author{Marcelo Salgado} 
\email{marcelo@nucleares.unam.mx}

\author{David Mart\'\i nez-del R\'\i o}
\email{david.martinez@nucleares.unam.mx}

\homepage{http://www.nucleares.unam.mx/~marcelo}
\affiliation{Instituto de Ciencias Nucleares 
\\ Universidad Nacional Aut\'onoma de M\'exico 
\\ Apdo. Postal 70--543 M\'exico 04510 D.F., M\'exico}

\date{\today}

\begin{abstract}
The initial value problem of scalar-tensor theories of gravity (STT) 
is analyzed in the physical (Jordan) frame using a 3+1 decomposition of 
spacetime. A first order strongly hyperbolic system is obtained for which the well posedness of the 
Cauchy problem can be established.
We provide two simple applications of the 3+1 system of equations: one for static and spherically symmetric 
spacetimes which allows the construction of unstable initial data (compact objects) 
for which a further black hole formation and scalar gravitational wave emission can 
be analyzed, and another application is for homogeneous and isotropic spacetimes that permits to study 
the dynamics of the Universe in the framework of STT.
\end{abstract}

\maketitle

\section{Introduction}
A physical meaningful theory is supposed to have a well posed Cauchy problem. Otherwise 
its predicting power during the Cauchy development can be lost.

In general it is not a trivial task to determine if a field theory possesses a well posed initial value problem. 
The systematic approach is to write the field equations as a manifestly hyperbolic (e.g. symmetric or strongly 
hyperbolic) system of partial 
differential equations (PDE). The existence and uniqueness theorems of such a system guarantee (under quite general hypothesis) 
that the solutions exist are unique and depend continuously on the initial data\footnote{If boundary conditions are taken into account 
continuous dependence upon these are to be considered as well.}.

In the case of gauge theories the analysis of the Cauchy problem becomes rather subtle or even complicated due to the gauge freedom 
of the theory. Indeed, the hyperbolicity of the PDE associated with gauge theories usually depends on the gauge choice. One must ensure that the chosen gauge is well 
behaved in the sense that it is preserved during the evolution if it holds in the initial spacelike hypersurface.

As regards general relativity (GR), it took several years and considerable effort to establish the well posedness of the 
Cauchy problem in the so-called {\it harmonic gauge} (e.g. see Refs. \cite{Choquet}). In fact this gauge allows to write the Einstein field equations as a set of quasilinear diagonal second order hyperbolic system for the metric components (the reduced field equations). The Bianchi identities ensure that the harmonic gauge is preserved 
during the Cauchy development if it holds initially.

During the last three decades a considerable work has been made for solving the Einstein's equations numerically. Unfortunately, 
several experiments have revealed that the harmonic gauge is not the best coordinate choice to evolve initial data that culminates in the 
formation of black holes. 

On the other hand, during the past few years, several novel formulations of Einstein's equations have been proposed 
for treating the Cauchy problem numerically. Unlike the traditional ones (notably the ADM formulation) these new formulations are strongly hyperbolic (see Ref. \cite{Reula,Alcubierre08} for a review). This analytic improvement together with new gauge conditions have dramatically enhanced the numerical stability during the 
evolution of several initial data. 
 
From the observational point of view there is little doubt about the validity of GR. So far almost all the predictions of the theory have been 
corroborated by experiments. It is expected that during the next decade or less the gravitational waves predicted by GR be directly 
detected by several interferometers. This information can be crucial for further testing the theory in the future.

Now, despite that GR seems to be the simplest diffeomorphic gravitational theory and so far the best one as observations are concerned, there are some astrophysical and cosmological data whose interpretation has renewed the interest for alternative theories of gravity. In the cosmological setting, 
the discovery of the accelerated expansion of the Universe together with the recent probes of the 
{\it cosmic microwave background}, points to the existence of some unknown 
type of matter (termed dark energy) which exerts negative pressure. The simplest implementation of this dark energy is to introduce a cosmological constant in the field equations. However, for several reasons (for instance, the so called coincidence problem) this is not considered to be  
the best solution to this enigma, and therefore cosmic dynamical scalar fields (termed generically as 
{\it quintessence}) have been proposed as alternatives 
for explaining such an acceleration with the property that their energy-density evolves with the scale factor. 

Very recently it has been advocated that {\it modified} theories of gravity can produce a cosmic 
acceleration without the introduction of a cosmological constant or of any other form of dark energy \cite{Modgrav}. These theories seem however to face serious 
problems at the solar system level \cite{Modgravprob} and show different kinds of instabilities \cite{Felice07,Seifert07}.

At smaller scales (galactic scales) the dynamics of stars and gas around spiral galaxies shows that their tangential velocity around the bulk of the 
galaxy is independent of the distance to the center. These observations contradict the Newtonian dynamics that was supposed to be valid at such 
regime. The explanation of this data (termed as {\it flat rotation curves of galaxies}, hereafter FRC) remains a mystery. The standard approach to solve this 
problem (as in cosmology) is to assume that another form of matter (dark matter) generates a galactic halo that interacts only gravitationally with the visible matter. This dark matter would produce an energy-density which decays with distance as $~1/r^\alpha$ (with $\alpha \approx 2$) such 
that the mass of the halo grows more less linearly with $r$ and therefore implies (within a Newtonian framework) a tangential velocity which is almost constant.

There exists also the view that the introduction of such dark matter does not really solve the problem in that the nature of such matter is still unknown. In this direction, the most popular and successful alternative explanation to the FRC without the need of 
any dark matter is the so-called {\it modified Newtonian dynamics} or MOND \cite{Mond}, which 
proposes that Newtonian dynamics is only valid for ``large" accelerations while for small accelerations the Newtonian gravity is compensated by 
a dynamics which is proportional to quadratic accelerations. Therefore, in the low acceleration regime, the modifications give rise to tangential velocities that are independent of the distance.

For many years the lack of a relativistic generalization for MOND was one of the main objections by its detractors. Very recently, Bekenstein suggested a 
tensor-vector-scalar theory of gravity (TeVeS) \cite{Bekenstein04} which is generally covariant and which reduces to the MOND model in some limit.
Needless to say, this theory should be considered as a viable alternative theory as long as it is compatible with 
all the observations (cosmology, solar system, binary pulsar, etc.) not only those related with the FRC. 
On the other hand, recent analyses indicate that TeVeS theory exhibits certain kind of instabilities too \cite{Seifert07} (see also Ref. \cite{Bruneton07} for several criticisms to TeVeS).

As mentioned in the beginning of this introduction, a serious candidate to replace GR should also have a well posed Cauchy problem. Therefore, in addition to the confrontation with observations, 
this is a task that should be tackled soon if some of these alternative theories are to be considered as serious theories.

Scalar-tensor theories of gravity (STT) are another alternative theories. 
The most prominent example of STT is perhaps the Brans-Dicke (BD) theory \cite{BD}. 
STT introduces a new fundamental scalar field which appears to be coupled non-minimally (NMC) to gravity (in the so-called Jordan frame). 
Due to this feature STT predict several phenomena not present in GR. STT are perhaps the best analyzed alternative theories of gravity. When the 
effective BD parameter is sufficiently large, the STT are compatible with the solar system tests. Moreover, 
STT have been proposed for explaining the acceleration of the Universe without any cosmological constant 
\cite{Boisseau00,dark,Schimd05}. 

Finally, from the mathematical point of view, STT posses a well posed Cauchy problem \cite{STTCauchy,Salgado06}. 

One of the main motivations of this contribution is to further discuss the Cauchy problem of STT in the Jordan 
frame \cite{Salgado06}. 
In fact, it is customary to find in the literature remarks about the unsuitability of the
Jordan frame to analyze the Cauchy problem \cite{Faraoni04}. This widespread prejudice is often invoked to transform the action (\ref{jordan}) (see the next section) to the so-called {\it conformal} or {\it Einstein} frame where 
the NMC function $F(\phi)$  is absorbed in the curvature by defining a non-physical conformal metric (the metric which 
is not associated to the non-null geodesics). The mathematical advantage of the conformal frame is that 
the resulting field equations, notably the equations for the non-physical metric, contain an effective energy-momentum 
tensor which does not include second order derivatives of the scalar field. This feature allows one to analyze the Cauchy problem 
of STT in the Einstein frame more or less in the same way as in GR, since basically the first order derivatives of the scalar-field 
which appears as ``sources" do not contribute to the principal part of the system of equations.
Of course the price to be payed when working within the conformal frame is that one is always dealing with non-physical fields 
(in the above sense) and that the transformation back to the physical frame has to be 
well defined. This is not always the case, in particular when $F(\phi)$ is not positive definite one 
can encounter situations where the conformal transformation is singular \cite{singular}
\footnote{In a very recent analysis 
\cite{Jarv07}, the authors argue that the connection between both frames also breaks in the limit of GR due to a lack of differentiability of the scalar-field redefinition in this limit. In fact the authors show that the 
cosmological attractor property of GR with respect to STT \cite{Damour93a} depends on the choice of the frame.}.

In this contribution we shall first remind the field equations of STT treated in 3+1 form \cite{Salgado06}.
Then we sketch the hyperbolicity analysis that leads to a well posed first order system of PDE.

In section 4, we consider the very simple case of initial data that represents a static and spherically symmetric spacetime. 
When such data correspond to an unstable configuration (e.g., a neutron star with maximal mass), the data can be used as a departure point for studying the gravitational collapse of compact objects within STT and the emission of scalar gravitational waves. A novel system of equations is presented using area-$r$ coordinates (spherical coordinates where the area 
of 2-spheres is given by $4\pi r^2$) and the system reduces to the one usually employed in GR in the absence of a scalar field. 

We provide in section 5, the fundamental equations required to construct an isotropic and homogeneous cosmological model for the 
Universe in the framework of STT.

\section{Scalar-tensor theories of gravity}

The general action for STT with a single scalar field is given by
\begin{equation}
  \label{jordan}
S[g_{ab}, \phi, \psi] = \int \left\{ \frac{F(\phi)}{16\pi G_0} R
-\left( \frac{1}{2}(\nabla \phi)^2 + V(\phi) \right) \right\} \sqrt{-g} d^4x
+ S_{\rm matt}[g_{ab}, \psi] \,\,\,\,,
\end{equation}
where $\psi$ represents collectively the matter fields (fields other
than $ \phi$; units where $c=1$ are employed) \footnote{Although the function $F(\phi)$ is, in principle,  completely arbitrary, 
it turns from the analysis of the Cauchy problem, that $F(\phi)>0$ is a necessary condition to avoid potential singularities 
or blow-ups in the equations and in several variables (this condition corresponds later 
to $f(\phi)>0$  in the notation adopted in this section).}.

The representation of STT given by Eq. (\ref{jordan}) is called the
{\it Jordan frame}. The covariant field equations obtained from variations of the action 
with respect to $g_{ab}$ and $\phi$ are:
\footnote{Latin indices from the first letters of the alphabet
$a,b,c,...$ are four-dimensional and run $0-3$.  Latin indices
starting from $i$ $(i,j,k,...)$ are three-dimensional and run $1-3$.}
\begin{eqnarray}  
\label{Einst}
G_{ab} &=& 8\pi G_0 T_{ab}\,\,\,\,, \\
\label{KGo}
\Box \phi &+& \frac{1}{2}f^\prime R = V^\prime \,\,\,,
\end{eqnarray}
where $^\prime$ indicates $\partial_\phi$, $G_{ab}=
R_{ab}-\frac{1}{2}g_{ab}R$ and
\begin{eqnarray}  
\label{effTmunu}
T_{ab} &:=& \frac{G_{{\rm eff}}}{G_0}\left(\rule{0mm}{0.5cm} T_{ab}^f + 
T_{ab}^{\phi} + T_{ab}^{{\rm matt}}\right)\,\,\,\,, \\
\label{TabF}
T_{ab}^f &:= & \nabla_a\left(f^\prime 
\nabla_b\phi\right) - g_{ab}\nabla_c \left(f^\prime 
\nabla^c \phi\right) \,\,\,\,, \\
T_{ab}^{\phi} &:= & (\nabla_a \phi)(\nabla_b \phi) - g_{ab}
\left[ \frac{1}{2}(\nabla \phi)^2 + V(\phi)\right ] \,\,\,\,, \\
\label{Geff}
G_{{\rm eff}} &:=& \frac{1}{8\pi f} \,\,\,\,,\,\,\,\,f:=\frac{F}{8\pi G_0} 
\,\,\,\,.
\end{eqnarray}

Using Eq. (\ref{Einst}), the Ricci scalar can be expressed in terms of
the effective energy-momentum tensor (EMT) Eq. (\ref{effTmunu}) and then
Eq. (\ref{KGo}) takes the following form,
\begin{widetext}
\begin{equation}
\label{KG}
{\Box \phi} = \frac{ f V^\prime - 2f^\prime V -\frac{1}{2}f^\prime
\left( 1 +  3f^{\prime\prime} 
\right)(\nabla \phi)^2 + \frac{1}{2}f^\prime T_{{\rm matt}} }
{f\left(1 + \frac{3{f^\prime}^2}{2f}\right) }\,\,\,\,,
\end{equation}
\end{widetext}
where $T_{{\rm matt}}$ stands for the trace of $T^{ab}_{{\rm matt}}$
and the subscript ``matt'' refers to the matter fields (fields other
that $\phi$).

Now, the Bianchi identities imply
\begin{equation}
\nabla _{c }T^{c a }=0\,\,\,\,.
\end{equation}
However, the use of the field equations leads to the conservation of the EMT of matter alone
\begin{equation}
\nabla _{c }T_{{\rm matt}}^{c a }=0\,\,\,\,,
\end{equation}
which implies the fulfillment of the (weak) equivalence principle. 

In which follows we shall use the 3+1 formalism of spacetime \cite{York79,Gour07} in order to recast the field 
equations as a Cauchy initial data problem.  

The definition of the 3+1 variables and the evolution equation obtained from these are respectively:
\begin{eqnarray}
\label{Q}
Q_a &:=& D_a \phi \,\,\,,\\
\label{Pidef}
\Pi &:=& {\cal L}_{\mbox{\boldmath{$n$}}} \phi \,\,\,,\\
\label{K_ab}
K_{ab}&:=& -\frac{1}{2} {\cal L}_{\mbox{\boldmath{$n$}}}\gamma_{ab} \,\,\,\,,\\
 {\cal L}_{\mbox{\boldmath{$n$}}} Q_a &=& \frac{1}{\alpha}D_a(\alpha\Pi) \,\,\,,
\end{eqnarray}
where ${\cal L}_{\mbox{\boldmath{$n$}}}$ is the Lie derivative
along $\mbox{\boldmath{$n$}}$ which is the unit time-like vector field normal to the spacelike hypersurfaces 
$\Sigma_t$ used to foliate the spacetime (as long as it is globally hyperbolic); $\mbox{\boldmath{$\gamma$}}$ is the 3-metric induced on $\Sigma_t$ and $D$ is the covariant derivative compatible with the 3-metric. 
In components $n^a= (1/\alpha,-\beta^i/\alpha)$ where $\alpha$ and 
$\beta^i$ are the {\it lapse} function and the {\it shift} vector, respectively, associated with the spacetime 
coordinates $(t,x^i)$.

Moreover, the 3+1 formalism allows one to write the field equations as initial data constraints and first order 
in time evolution equations (see Ref. \cite{Salgado06})
\footnote{A slightly different notation 
with respect to \cite{Salgado06} has been employed here in order to match with the notation used in many references on numerical relativity. In order to return to the notation of Ref. \cite{Salgado06}, one needs to perform the transformation 
$\alpha\rightarrow N$, $\beta^ i\rightarrow -N^i$, $\gamma_{ij}\rightarrow h_{ij}$ for the lapse, shift and the 3-metric 
respectively.}. 
The field equations (\ref{Einst}) then read as the following constraints (\ref{CEHfSST}) and (\ref{CEMfSST}) 
and the evolution equation (\ref{EDEfSST}):
\begin{equation}  
\label{CEHfSST}
^3 R + K^2 - K_{ij} K^{ij} 
- \frac{2}{f}\left[
 f^\prime \left(\rule{0mm}{0.4cm} D_l Q^l + 
K \Pi \right) + \frac{\Pi^2}{2} + \frac{Q^2}{2}
\left(\rule{0mm}{0.4cm} 1 + 2f^{\prime\prime}\right) \right]
  = \frac{2}{f} \left[\rule{0mm}{0.4cm} E_{\rm matt} + V(\phi)\right]  \,\,\,,
\end{equation}

\begin{equation}  
\label{CEMfSST}
 D_l K_{\,\,\,\,\,i}^{l} - D_i K
+\frac{1}{f}\left[\rule{0mm}{0.5cm}
f^\prime\left(\rule{0mm}{0.4cm} K_{i}^{\,\,l} Q_l + \,D_i\Pi
\right) + \Pi Q_i\left(\rule{0mm}{0.4cm} 1 + f^{\prime\prime}\right)\right]
= \frac{1}{f} J_i^{\rm matt} \,\,\,\,,
\end{equation}

\begin{eqnarray}  
\label{EDEfSST}
& & \partial_t K_{\,\,\,j}^i - \beta^l \partial_l K_{\,\,\,j}^i - K_{\,\,\,l}^i
\partial_j \beta^l + K_{\,\,\,j}^l \partial_l \beta^i 
+ D^i D_j \alpha  -\,^3 R_{\,\,\,j}^i \alpha - \alpha K K_{\,\,\,j}^i
\nonumber \\ 
&&  
+ \frac{\alpha}{f}\left[\rule{0mm}{0.6cm} Q^i Q_j\left(\rule{0mm}{0.4cm} 1
+ f^{\prime\prime}\right)  
+ f^\prime \left(\rule{0mm}{0.4cm} D^i Q_j
+ \Pi K^{i}_{\,\,\,j} \right) \right]  
- \frac{\delta_{\,\,\,j}^i \alpha}{2f\left(1 + \frac{3{f^\prime}^2}{2f}\right)}
\left(\rule{0mm}{0.6cm} Q^2-\Pi^2\right)\left( \frac{{f^\prime}^2}{2f}
- f^{\prime\prime}\right) 
\nonumber \\
&& = -\frac{\alpha}{2f\left(1 + \frac{3{f^\prime}^2}{2f}\right)}
\left\{ 2 S_{{\rm matt}\,\,\,j}^i \left(1 + \frac{3{f^\prime}^2}{2f} \right)
+ \delta_{\,\,\,j}^i \left[ f^\prime V^\prime
+ 2V\left(1 + \frac{{f^\prime}^2}{2f} 
\right) - \left(\rule{0mm}{0.4cm} S_{\rm matt}
- E_{\rm matt}\right) \left(1 + \frac{{f^\prime}^2}{f} \right)
\right]\right\}\,,
\end{eqnarray}
where $E_{\rm matt}:= n^a n^b T_{ab}^{{\rm
matt}}$, $S^{ab }_{\rm matt}:=\,\gamma_{\,\,\,c}^a \gamma_{\,\,\,d}^b
T^{cd}_{\rm matt}$, $S_{\rm matt}:=S^{a}_{\,\,a\,\,\,\rm matt}$, $J^a_{\rm matt} 
:= -n_d \gamma_{\,\,\,c}^a T^{cd}_{\rm matt}$ 
and  $\gamma_{\,\,\,c}^a= \delta_{\,\,\,c}^a + n^a n_c$ is the projector onto $\Sigma_t$. 

From Eqs. (\ref{CEHfSST}) and (\ref{EDEfSST}), a useful evolution equation for the trace of the
extrinsic curvature can be obtained:
\begin{eqnarray}  
\label{EDKSTT}
&& \partial_t K -\beta^l \partial_l K + \,^3\Delta \alpha
- \alpha K_{ij} K^{ij} 
- \frac{\alpha f^\prime}{f} \left(\rule{0mm}{0.4cm} D_l Q^l
+ \Pi K \right) \nonumber \\
&& 
-\frac{\alpha}{f\left(1 + \frac{3{f^\prime}^2}{2f}\right)}
\left\{ \Pi^2 \left( 1 + \frac{3{f^\prime}^2}{4f}
+ \frac{3f^{\prime\prime}}{2} \right)
+ Q^2\left[ \frac{3{f^\prime}^2}{4f}\left(\rule{0mm}{0.4cm} 1
+ 2 f^{\prime\prime}\right)  
- \frac{f^{\prime\prime}}{2}\right]  \right\} \nonumber \\
&& = \frac{\alpha}{2f\left(1 + \frac{3{f^\prime}^2}{2f}\right)}
\left\{ S_{\rm matt} + E_{\rm matt} \left(1 + \frac{3{f^\prime}^2}{f}\right) 
- 2V\left(1- \frac{3{f^\prime}^2}{2f}\right) - 3 f^\prime V^\prime \right\}
 \,\,\,,
\end{eqnarray}
where $\,^3\Delta$ is the Laplacian compatible with the 3-metric $\gamma_{ij}$, and $Q^2= Q_l Q^l$.
\bigskip

The evolution equation for $\Pi$ is obtained from Eq. (\ref{KG}) and 
reads as follows \cite{Salgado06},
\begin{equation}
\label{evPi1}
{\cal L}_{\mbox{\boldmath{$n$}}} \Pi  - \Pi K - Q^c D_c[{\rm ln}\alpha] 
- D_c Q^c  
= - \frac{ f V^\prime - 2f^\prime V -\frac{1}{2}f^\prime
\left( 1 +  3f^{\prime\prime} 
\right)\left(\rule{0mm}{0.4cm} Q^2-\Pi^2\right)
+ \frac{1}{2}f^\prime T_{{\rm matt}} }
{f\left(1 + \frac{3{f^\prime}^2}{2f}\right) } \,\,\,,
\end{equation}
where $T_{{\rm matt}}= S_{{\rm matt}}-E_{{\rm matt}}$.
\bigskip

\bigskip

\section{Hyperbolic system of equations}
The ``gravitational" sector of the equations (\ref{CEHfSST})$-$(\ref{EDKSTT}) (i.e., the parts that remain in the absence of the scalar field) corresponds to the 
usual ADM equations {\it \`a la} York \cite{York79} (hereafter ADMY system). These equations do not form a strongly hyperbolic system, but only a weakly hyperbolic one,  for which the Cauchy problem is not well posed . Therefore, the inclusion of the non-minimally coupled 
scalar field will not improve the situation.

In pure GR various strongly hyperbolic systems can be constructed from the ADMY equations 
\cite{Alcubierre08}. The approach consists in recasting all the equations in full first order form (in space and time), and then to add appropriately multiples of the constraints to 
the evolution equations. In this way one can obtain a system in the form
\begin{equation}
\label{PDE}
\partial_t {\vec u} + \mathbb{M}^i \nabla_i {\vec u} = {\vec S}({\vec u}) \,\,\,,
\end{equation}
where ${\vec u}$ represents collectively the fundamental variables (like the 
$h_{ij}$'s, $K_{ij}$'s, etc.), $\mathbb{M}^i$ is called the 
{\it characteristic} matrix of the system (which depends on ${\vec u}$ but is independent of its derivatives)
and ${\vec S}({\vec u})$ are {\it source} terms which include only the fundamental variables (not their derivatives).

The addition of the constraints can make the characteristic matrix to be diagonalizable (i.e., the matrix 
has a real set of eigenvalues and a complete set of eigenvectors). 

Although there are linear-algebraic methods (see for instance \cite{KST}) to prove the strongly hyperbolic character of the 
characteristic matrix, one can also follow another economical but equivalent approach. This consists in constructing the 
eigenfields ${\vec w}:= \mathbb{R}^{-1} {\vec u}$ from the eigenvector matrix $\mathbb{R}$ with the property 
$\mathbb{L} = \mathbb{R}^{-1} \mathbb{M}^x \mathbb{R}$ is the diagonal eigenvalue matrix associated with 
a particular direction `$x$' of propagation. The point is that one can achieve such a construction by a {\it judicious guessing} 
method (see the Appendix for a simple example). This is roughly speaking the approach followed by Bona-Mass\'o \cite{Bona92}.

Of course, another key element in the construction of a strongly hyperbolic system is the gauge choice. In our case we consider 
the shift as a given function of the coordinates and a modified Bona-Mass\'o time slicing which reads
\begin{equation} 
\label{STTBMlapse}
\partial_0 \alpha   = - f_{\rm BM} (\alpha)\alpha^2  
\left(\rule{0mm}{0.4cm} K- \frac{\Theta}{f_{\rm BM}}\frac{f^\prime}{f} \Pi\right) \,\,\,\,,
\end{equation}
where $\partial_0= \partial_t -\beta^l\partial_l$, $f_{\rm BM}(\alpha) > 0$ (i.e., $f_{\rm BM}$ is a positive but 
otherwise arbitrary function of the lapse) and $\Theta$ is a parameter \footnote{This slicing
condition is a bit modification with respect to the one considered in
\cite{Salgado06} in that the $\Theta$ term contains the factor $f_{\rm BM}$ in
the denominator. However, both conditions agree when adopting the 
modified harmonic condition (termed pseudo harmonic in \cite{Salgado06}) $\Theta=1=f_{\rm BM}$.}. 

In pure GR the usual BM condition [i.e., Eq. (\ref{STTBMlapse}) with $\Theta\equiv 0$] 
generalizes the harmonic slicing $f_{BM}=1$. The BM condition therefore includes several choices of 
$f_{\rm BM}(\alpha)$ that can be more useful than others depending on the evolution to be analyzed 
(see \cite{Alcubierre08} for a more detailed discussion). However, the usual BM condition 
seems not to work for STT in the Jordan frame. In fact it is possible 
to show \cite{Martinez07} that taking $\Theta\equiv 0$ leads to a weakly hyperbolic 
system. Therefore, the modification of the usual BM slicing condition is a crucial step 
towards a strongly hyperbolic system of STT when formulated in the Jordan frame.

In this way, the evolution system in terms of the eigenfields and their corresponding eigenvalues reads \cite{Martinez07}
\begin{equation} 
\label{Hyp}
\partial_0 w^i+  \partial_x \left(\lambda^i  w^i\right) \simeq 0
\end{equation}
(here no sum over the repeated index) where the symbol $\simeq$ indicates {\it equal up to 
principal part}; $w^i$ represents 34 eigenfields constructed upon the first order variables defined in Sec.II, and also 
from first order variables defined in terms of the spatial derivatives of the 3-metric and the spatial derivatives of the lapse. In pure GR there are only 30 eigenfields  (see Ref. \cite{Alcubierre08} for a review) since $Q_i\equiv 0\equiv \Pi$. 
Here 
$\lambda^i$ is the eigenvalue which is associated with the speed of propagation of the corresponding eigenfield $w^i$. In fact there are 20 eigenfields which evolve along the time lines (with speed $-\beta^x$ 
while the remaining 14 propagate with non trivial speeds some of which depend on the gauge function $f_{BM}(\alpha)$. Again, 
it is important to emphasize that the strong hyperbolicity arises only with the value $\Theta \neq 0$, otherwise it is not possible 
to construct all the eigenfields. Moreover, with the optimal value $\Theta =1$, together with 
a positive definite NMC function $f(\phi)$, the eigenfields are all smooth in their variables. 
We remark $\Theta =1$ is also the required optimal value in the second order approach 
(Cf. Ref. \cite{Salgado06}) for the reduced system of field equations 
to be in quasilinear diagonal second order hyperbolic form ($f_{BM}=1$ was assumed in Ref. \cite{Salgado06},  and the shift was also included as a {\it live shift}).

Basically both the second and the first order analysis sketched here show that STT posses a well posed Cauchy problem 
in the Jordan frame. 

\section{Static and spherically symmetric initial data}
In this section we shall be concerned with initial data that represent a static and spherically symmetric (SSS) spacetime. 
In fact, we will be interested mainly in initial data corresponding to static but {\it unstable} configurations of compact objects (namely, neutron or boson stars). Here by unstable 
we mean a configuration that is near to the transition to 
spontaneous scalarization \cite{Damour93,Salgado98,Novak,Whinnett00} or near to the point (i.e., maximal mass) of 
gravitational collapse towards a black hole. In this way, a small perturbation to the unstable configuration will lead 
to a non trivial dynamics that can be accompanied by the emission of scalar gravitational waves \cite{Novak}.

First, staticity implies that the extrinsic curvature and the shift vector will be globally null. Moreover, the field variables 
will be time-independent. In this case the l.h.s of the momentum constraint (\ref{CEMfSST}) becomes identically null which implies $J_i^{\rm matt}\equiv 0$. The remaining field equations read as follows:
\begin{equation}  
\label{CEHfSSTstat}
^3 R - \frac{2}{f}\left[
 f^\prime D_l Q^l  + \frac{Q^2}{2}
\left(\rule{0mm}{0.4cm} 1 + 2f^{\prime\prime}\right) \right]
  = \frac{2}{f} \left[\rule{0mm}{0.4cm} E_{\rm matt} + V(\phi)\right]  \,\,\,,
\end{equation}

\begin{eqnarray}  
\label{EDEfSSTstat}
 && D^i D_j \alpha  -\,^3 R_{\,\,\,j}^i \alpha 
+ \frac{\alpha}{f}\left[\rule{0mm}{0.6cm} Q^i Q_j\left(\rule{0mm}{0.4cm} 1
+ f^{\prime\prime}\right)  
+ f^\prime  D^i Q_j \right]  
- \frac{\alpha \delta_{\,\,\,j}^i  Q^2\left( \frac{{f^\prime}^2}{2f}
- f^{\prime\prime}\right) }{2f\left(1 + \frac{3{f^\prime}^2}{2f}\right)} 
\nonumber \\
&& = -\frac{\alpha}{2f\left(1 + \frac{3{f^\prime}^2}{2f}\right)}
\left\{ 2 S_{{\rm matt}\,\,\,j}^i \left(1 + \frac{3{f^\prime}^2}{2f} \right)
+ \delta_{\,\,\,j}^i \left[ f^\prime V^\prime
+ 2V\left(1 + \frac{{f^\prime}^2}{2f} 
\right) - \left(\rule{0mm}{0.4cm} S_{\rm matt}
- E_{\rm matt}\right) \left(1 + \frac{{f^\prime}^2}{f} \right)
\right]\right\}\,,
\end{eqnarray}
\begin{eqnarray}  
\label{EDKSTTstat}
&& \,^3\Delta \alpha
- \frac{\alpha f^\prime}{f}  D_l Q^l 
-\frac{\alpha}{f\left(1 + \frac{3{f^\prime}^2}{2f}\right)}
\frac{Q^2}{2}\left[ \frac{3{f^\prime}^2}{2f}\left(\rule{0mm}{0.4cm} 1
+ 2 f^{\prime\prime}\right)  
- f^{\prime\prime}\right]  \nonumber \\
&& = \frac{\alpha}{2f\left(1 + \frac{3{f^\prime}^2}{2f}\right)}
\left\{ S_{\rm matt} + E_{\rm matt} \left(1 + \frac{3{f^\prime}^2}{f}\right) 
- 2V\left(1- \frac{3{f^\prime}^2}{2f}\right) - 3 f^\prime V^\prime \right\}
 \,\,\,,
\end{eqnarray}
\begin{equation}
\label{evPi1stat}
Q^l D_l[{\rm ln}\alpha] + D_l Q^l  
=  \frac{ f V^\prime - 2f^\prime V -\frac{1}{2}f^\prime
\left( 1 +  3f^{\prime\prime} 
\right) Q^2
+ \frac{1}{2}f^\prime T_{{\rm matt}} }
{f\left(1 + \frac{3{f^\prime}^2}{2f}\right) } \,\,\,.
\end{equation}

Moreover, using Eq. (\ref{evPi1stat}) one can replace the terms with $D_l Q^l$ in Eqs. (\ref{CEHfSSTstat}) and (\ref{EDKSTTstat}) to obtain respectively
\begin{eqnarray}  
\label{CEHfSSTstat2}
^3 R = && \frac{2}{f \left(1 + \frac{3{f^\prime}^2}{2f}\right)} \left[\rule{0mm}{0.4cm} -f^\prime \left(1 + \frac{3{f^\prime}^2}{2f}\right)\frac{Q^l}{\alpha} D_l\alpha +
\frac{Q^2}{2}\left(1+\frac{{f^\prime}^2}{2f}+ 2f^{\prime\prime}\right)\right. \nonumber \\
&& \left. \rule{0mm}{0.4cm}
+ V(\phi)\left(1-\frac{{f^\prime}^2}{2f}\right) + 
f^\prime V^\prime + \frac{{f^\prime}^2}{2f}S_{\rm matt} + \left(1+\frac{{f^\prime}^2}{f}\right)E_{\rm matt}  \right]= 16\pi G_0 
E  \,\,\,,
\end{eqnarray}
\begin{eqnarray}  
\label{EDKSTTstat2}
 \,^3\Delta \alpha
+ \frac{f^\prime}{f} Q^l  D_l \alpha = && \frac{\alpha}{2f\left(1 + \frac{3{f^\prime}^2}{2f}\right)}
\left\{ Q^2\left(f^{\prime\prime}+ \frac{{f^\prime}^2}{2f}\right)  
- 2 f^\prime V^\prime - 2V\left(1+ \frac{{f^\prime}^2}{2f}\right)\right. \nonumber \\
&& \left. +\left(1+\frac{{f^\prime}^2}{f}\right)S_{\rm matt}
+\left(1-\frac{{f^\prime}^2}{f}\right)E_{\rm matt}\right\}
 \,\,\,.
\end{eqnarray}
where $E=n^a n^b T_{ab}$ is the effective energy-density which incorporates all the energy-density contributions 
(including the scalar field and the NMC). For instance, in the minimal coupling case 
$f=1/(8\pi G_0)={\rm const.}$, therefore $E= E_{\rm matt}+ \frac{Q^2}{2}+V(\phi)$ which corresponds to the energy-density of an ordinary 
scalar field plus the energy density of matter.

Equations (\ref{EDKSTTstat2}) and (\ref{evPi1stat}) are 
elliptic equations for the lapse and the scalar field respectively 
(notice $D_l Q^l = \,^3\Delta \phi$).

\bigskip

Finally, the conservation equation
\begin{equation}
\label{Eqmatt}
\nabla _{c }T_{{\rm matt}}^{c a }=0\,\,\,\,,
\end{equation}
together with the staticity assumption will provide the {\it equilibrium} equation for the matter which we have not specified yet.

\subsection{Spherical symmetry}
One of the simplest applications of the field equations of STT is for SSS 
spacetimes, which can be described by the following metric in area-$r$ coordinates 
\begin{equation}
\label{metric}
ds^2=  -\alpha^2(r) dt^2 + A^2(r) dr^2 + r^2d\theta^2 
+ r^2\sin^2(\theta)\, d\varphi^2\,\,\,.
\end{equation}
Now, adopting the parametrization $A^2(r)= \left( 1 - \frac{2G_0 m(r)}{r}\right)^{-1}$, 
one can easily verify that Eq. (\ref{CEHfSSTstat2}) reduces to
\begin{equation}
\label{Eqm}
\partial_r m = 4\pi r^2 E\,\,\,,
\end{equation}
where $E$ is to be read-off from Eq. (\ref{CEHfSSTstat2}) with $\frac{Q^l}{\alpha} D_l\alpha= 
(Q_r \partial_r \alpha)/(\alpha A^2)$ and $Q^2= (Q_r)^2/A^2$. The integration of Eq. (\ref{Eqm}) provides the ``mass function" 
$m(r)$ which at infinity (and for asymptotically flat spacetimes) corresponds to the ADM-mass of the configuration.

For the lapse we can use Eq. (\ref{EDKSTTstat2}). However, it turns that the $\theta-\theta$ component of Eq. (\ref{EDEfSSTstat}) 
provides a first order equation for the lapse. In order to arrive to that equation we use 
$^3 R_{\,\,\,\theta}^\theta= G_0m/r^3 + (G_0\partial_r m)/r^2= G_0m/r^3 + 4\pi G_0 E$. Then one can verify that 
Eqs. (\ref{EDEfSSTstat}) and (\ref{CEHfSSTstat2})
together with the explicit expression for $E$ lead to
\begin{equation}
\label{Eqlapse}
\frac{\partial_r \alpha}{\alpha}= \frac{A^2}{1+\frac{rf^\prime}{2f}Q_r}\left\{\rule{0mm}{0.5cm}  \frac{G_0 m}{r^2} + \frac{r}{2f}\left[
S_{{\rm matt}\,\,\,r}^r - V(\phi) + \frac{(Q_r)^2}{2A^2} - \frac{2f^\prime}{rA^2} Q_r\right]\right\} \,\,\,.
\end{equation}
Finally Eqs. (\ref{evPi1stat}) and (\ref{Q}) read respectively
\begin{equation}
\label{EqQ}
\partial_r Q_r + Q_r \left[ \frac{2}{r} + \frac{\partial_r \alpha}{\alpha}- A^2\left(4\pi G_0 r E - \frac{G_0 m}{r^2}\right)\right]
= A^2 \frac{ f V^\prime - 2f^\prime V -\frac{1}{2}f^\prime
\left( 1 +  3f^{\prime\prime} 
\right) Q^2
+ \frac{1}{2}f^\prime T_{{\rm matt}} }
{f\left(1 + \frac{3{f^\prime}^2}{2f}\right) } \,\,\,,
\end{equation}
\begin{equation}
\label{Qr}
\partial_r \phi = Q_r  \,\,\,.
\end{equation}
Equations (\ref{Eqm})$-$(\ref{Qr}) together with Eq. (\ref{Eqmatt}) are the set of fundamental equations 
to be solved for a SSS spacetime in the area-$r$ coordinates. Notice that when one replaces 
the term $\partial_r \alpha/\alpha$ appearing in $E$ and in Eq. (\ref{EqQ}) by the r.h.s of Eq. (\ref{Eqlapse}), 
the fundamental equations have all the form $\partial_r \vec{U} = \vec{\mathscr{H}}(r,\vec{U})$ where $\vec{U}$ stands for the field variables $(\phi,Q_r,{\rm ln}\alpha,m)$,
and $\vec{\mathscr{H}}$ are non linear functions of their arguments but independent of the derivatives of the fields $\vec{U}$. One can therefore employ a Runge-Kutta algorithm to solve the equations numerically. As far as we are aware this system of equations has not been considered before, except for very particular cases. 
In fact for the specific class of STT where $f=(1+ 16\pi G_0 \xi \phi^2)/(8\pi G_0)$ and 
$T_{{\rm matt}}^{ab} = (\rho + p)u^a u^b + g^{ab} p$ is the EMT of a perfect fluid, Eqs. (\ref{Eqm})$-$(\ref{Qr}) 
reduce to those of Ref. \cite{Salgado98} where the phenomenon of spontaneous scalarization \cite{Damour93} was further analyzed. 
For a perfect fluid Eq. (\ref{Eqmatt}) yields the equation of hydrostatic equilibrium 
\begin{equation}
\label{hydro}
\partial_r p = -(\rho + p) \frac{\partial_r \alpha}{\alpha} \,\,\,.
\end{equation}
Moreover, $S_{{\rm matt}\,\,\,r}^r= p$, $E_{\rm matt}= \rho$, $T_{{\rm matt}}= 3p-\rho$. This differential equation can also 
be written in terms 
of the baryon density $n(r)$ since for neutron stars the equation of state is usually parametrized as $p=p(n)$, $\rho=\rho(n)$. We have
\begin{equation}
\label{hydro2}
\partial_r n = -\frac{(\rho + p) n}{\Gamma p} \frac{\partial_r \alpha}{\alpha} \,\,\,,
\end{equation}
where $\Gamma:= d {\rm ln[p]}/d {\rm ln[n]}$ is the adiabatic index. Then $n(r)$ adds to $\vec{U}$ as a further field.

As one can easily observe, in the absence of a scalar field [i.e.,$f=1/(8\pi G_0)={\rm const.}$, $\phi\equiv 0
\equiv V(\phi) \equiv Q_r $], 
Eqs. (\ref{Eqm}) and (\ref{Eqlapse}) with (\ref{hydro}) or (\ref{hydro2}) are the equations of hydrostatic equilibrium for a relativistic star in pure GR.
\bigskip

The system of equations presented in this section can be used to construct 
unstable SSS initial data (compact objects) that can serve to trigger the dynamical transition to 
spontaneous scalarization or to trigger the gravitational collapse of an scalarized object towards a black hole formation 
(cf. Ref. \cite{Novak}). One expects that if the static configuration is very close to the point of instability, a small perturbation to the unstable static solution will drive the configuration to the final stage which can be a scalarized object (if the initial 
configuration is an unscalarized neutron star near to the point to the phase transition) or a black hole (if 
the initial data represent a scalarized neutron star with maximal mass). 

Another possibility is to use a boson field as a matter model \cite{Whinnett00}. In that case, the equation of hydrostatic equilibrium (\ref{hydro}) is 
replaced by the Klein-Gordon equation associated with the boson field. 

\section{Isotropic and Homogeneous Spacetimes}
Another simple application of the 3+1 system of equations of Sec. II is for a Friedmann-Robertson-Walker (FRW) cosmology where the metric is given by
\begin{equation}
\label{FRWmetric}
ds^2=  - dt^2 + a^2(t) \left[ \frac{dr^2}{1-kr^2} + r^2d\theta^2 
+ r^2\sin^2(\theta)\, d\varphi^2\right] \,\,\,,
\end{equation}
which describes an isotropic and homogeneous spacetime. In this case the 
extrinsic curvature is given by $K^{i}_{\,\,\,j}= -\frac{\dot a}{a} \delta^{i}_{\,\,j}$, and so $K= -3\frac{\dot a}{a}
\equiv -3 H(t)$, where $H(t)$ is the Hubble expansion rate, and $^3 R= 6k/a^2$ (where $k=\pm 1,0$). Here a dot indicates derivative with respect to cosmic time $t$. The Hamiltonian constraint (\ref{CEHfSST}) reduces to the Friedmann equation
\begin{equation}
\label{Fried}
H^2 + \frac{k}{a^2}= \frac{1}{3f} \left[\rule{0mm}{0.4cm} E_{\rm matt} + V(\phi) + \frac{\Pi^2}{2} - 3f^\prime H \Pi \right]\,\,\,.
\end{equation}

On the other hand Eq. (\ref{EDKSTT}) provides the dynamic equation that governs the acceleration of the Universe 
\begin{equation}
\label{accel}
\dot H + H^2
- \frac{f^\prime}{f} \Pi H 
= -\frac{1}{6f\left(1 + \frac{3{f^\prime}^2}{2f}\right)}
\left\{ 2\Pi^2 \left( 1 + \frac{3{f^\prime}^2}{4f}
+ \frac{3f^{\prime\prime}}{2} \right) +
S_{\rm matt} + E_{\rm matt} \left(1 + \frac{3{f^\prime}^2}{f}\right) 
- 2V\left(1- \frac{3{f^\prime}^2}{2f}\right) - 3 f^\prime V^\prime \right\}
 \,\,\,,
\end{equation}

Finally, Eqs. (\ref{Pidef}) and (\ref{evPi1}) take respectively the form
\begin{equation}
\dot \phi = \Pi \,\,\,,
\end{equation}

\begin{equation}
\label{evPicosmo}
\dot \Pi  + 3 \Pi H 
= - \frac{ f V^\prime - 2f^\prime V +\frac{1}{2}f^\prime
\left( 1 +  3f^{\prime\prime} 
\right) \Pi^2
+ \frac{1}{2}f^\prime T_{{\rm matt}} }
{f\left(1 + \frac{3{f^\prime}^2}{2f}\right) } \,\,\,.
\end{equation}

Eqs. (\ref{Fried})$-$(\ref{evPicosmo}) together with $H= \dot a/a$ are the fundamental equations that determine the dynamics of a 
homogeneous and isotropic Universe. Like in previous section, we emphasize that in the absence of a scalar field 
Eqs. (\ref{Fried}) and (\ref{accel}) reduce to the very well known equations in GR.

Assuming a perfect fluid for the matter we have again 
$S_{\rm matt}= 3p$, $E_{\rm matt}= \rho$, $T_{{\rm matt}}= 3p-\rho$. The conservation Eq. (\ref{Eqmatt}) then yields
\begin{equation}
\dot \rho + 3\left( \rho + p\right) H =0 \,\,\,,
\end{equation}
which for a non interacting two fluid component (photons and pressureless matter) provides the matter energy-density as a function of the scale factor: $\rho= c_1/a^4 + c_2/a^3$ where $c_1$ and $c_2$ are 
constants that are fixed, for instance, with the values of the photon and matter densities today.

As in the previous section, the cosmological equations (\ref{accel})$-$(\ref{evPicosmo})
have all the form
$\dot{\vec{U}} = \vec{\mathscr{H}}(t,\vec{U})$ \footnote{Actually the cosmic time does not appear explicitly in the 
equations. However, it is 
often convenient to use $\tau:= {\rm ln}[a(t)/a_0]$ (where $a_0$ is the scale factor today) as ``time" variable instead of $t$. Then the equations take the form $\dot{\vec{U}} = \vec{\mathscr{H}}(\tau,\vec{U})$, where now 
$\vec{U}$ stands for $(\phi,\Pi,H)$ and the dot indicates derivative with respect to $\tau$. In this case, the explicit 
dependence of $\vec{\mathscr{H}}(\tau,\vec{U})$ with respect to $\tau$ arises because the matter energy-density, $\rho$, depends explicitly on the scale factor $a$.} where now $\vec{U}$ stands for the field variables $(\phi,\Pi,a,H)$,
and $\vec{\mathscr{H}}$ are non linear functions of their arguments but independent of the time derivatives of the fields $\vec{U}$. 
The form of the equations again permits the use of a Runge-Kutta method to solve the system numerically. 
In this respect, when the system is evolved backwards in time 
by taking as initial data the conditions of the Universe today, the Hamiltonian 
constraint 
(\ref{Fried}) can be used to check the numerical accuracy at every time step. 
In particular, the initial data cannot be arbitrary but has to be consistent with 
Eq. (\ref{Fried}) as well.

Indeed for the particular class of STT given by 
$f=(1+ 16\pi G_0 \xi \phi^2)/(8\pi G_0)$ a similar system of equations was used in \cite{Gmodel} to analyze in detail 
several cosmological models which can be viewed as models of the Universe with an effective gravitational ``constant" 
that evolves in cosmic time. 

For the case with an arbitrary $f(\phi)$, equations similar to the system (\ref{Fried})$-$(\ref{evPicosmo}) 
have been analyzed in the literature \cite{Boisseau00,Schimd05} although not obtained from a 3+1 formalism and therefore 
they are presented in a more complicated fashion.
Using the Brans-Dicke parametrization, an analogous system of equations can be seen in Ref. \cite{Faraoni04}.

\section{Outlook and Conclusions}
We have presented the scalar tensor theories of gravity in the Jordan frame and wrote 
their corresponding 3+1 equations (constraint equations and evolution equations) in order to analyze the Cauchy initial value problem. Then we briefly discussed the 
procedure that led to a set of (strongly) hyperbolic equations for which a well posed Cauchy problem can be established. This analysis shows that the STT in the Jordan frame have a well posed Cauchy problem, contrary to what it is usually claimed 
in the literature.
Two simple applications of the equations were presented. First the case of static and spherically symmetric spacetimes where  
a set of ``master'' equations were obtained for arbitrary NMC functions $f(\phi)$. These equations are specially suited for a numerical analysis.
 For a specific class of STT such 
equations have in fact showed their usefulness \cite{Salgado98}. 
This situation is very interesting for the study of the phenomenon of spontaneous scalarization that arises in compact objects. 
Moreover, configurations of this kind at the threshold of instability (towards spontaneous scalarization or towards the 
collapse into a black hole) can serve as a physical initial data for studying the dynamics of such objects in STT. 
For instance, recent studies show that spontaneous scalarization can have an impact on the signature of the wave forms emitted by 
a perturbed neutron star \cite{Kokkotas06}. Such wave forms when compared with future observations can be useful to validate some classes of STT or 
to bound the non-minimal coupling constants. 
Finally, we discussed the case of isotropic and homogeneous spacetimes (FRW cosmology) in STT. We again obtained a set of fundamental equations which govern the dynamics of the Universe in the context of STT.  These equations have been quite 
useful for the numerical analysis of cosmologies with specific choices of $f(\phi)$ \cite{Gmodel}.

In the near future we plan to use the formalism presented here to study the dynamics and transition to spontaneous 
scalarization in different kinds of compact objects. Moreover, in the cosmological setting we expect to tackle the acceleration 
of the Universe and the dark energy problem without the introduction of a cosmological constant but in the framework of STT. This issue has been extensively analyzed in the past (see Refs. \cite{Faraoni04,Maeda03} for a review), 
nevertheless we consider that the formalism presented here allows to appreciate the NMC contributions in a more 
transparent fashion. 

\begin{acknowledgments}
We thank M.Alcubierre and D.Nu\~nez for discussions and comments. This work has been partially supported by DGAPA-UNAM, grant No. IN119005.
\end{acknowledgments}

\bigskip
\appendix
\section{Simple example of an hyperbolic system}
Let us consider the system
\begin{eqnarray}
\label{toy1}
 \partial_t \Pi + a\,\partial_x \Pi + b\,\partial_x Q = S_\Pi(Q,\Pi) \,\,\,,\\
\label{toy2}
 \partial_t Q + c\,\partial_x \Pi + d\,\partial_x Q = S_Q (Q,\Pi) \,\,\,\,,
\end{eqnarray}
where the coefficients $a-d$ are in general functions of $(t,x)$ and $S_{\Pi,Q}(Q,\Pi)$ 
are {\it source} functions.

Using the {\it judicious guessing} approach we shall construct a new system of the form (\ref{Hyp}) which is 
manifestly hyperbolic and then we shall confront this method with the more systematic method which uses linear algebra. 

First we ask for a linear combination $w:=\Pi + \sigma Q$ to be the eigenfunction. This means that we write an evolution equation 
for $w$, where $\sigma(a,b,c,d)$ is a function of the coefficients $a-d$ which is to be found 
such that the r.h.s of the evolving part take the form $\lambda_{\sigma} \, \partial_x\left( \Pi + \sigma Q \right) $, where $\lambda_{\sigma}$ will be the eigenvalue(s). 
In this way we have

\begin{eqnarray}
\label{jg1}
&\partial_t \left(\Pi + \sigma Q\right) &\simeq - \partial_x \left[\rule{0mm}{0.4cm} \left( a+\sigma c\right) \Pi + \left( b + \sigma d\right) Q \right] \nonumber \\
& &\simeq -\left( a+\sigma c\right) \,\partial_x\left(  \Pi + \frac{b+\sigma d}{a+\sigma c} Q\right) \,\,\,\,.
\end{eqnarray}
Note that with the symbol $\simeq$ we have discarded from the analysis all the terms which do not contain derivatives of 
$\Pi$ and $Q$ (including the sources). In this sense we can consider as though the coefficients $a-d$ were constants. So, by comparing
the coefficients of $Q$ in the evolving $\partial_t$ and the spatial $\partial_x$ parts of Eq. (\ref{jg1}), and then imposing 
$\sigma= (b+\sigma d)/(a+\sigma c)$, one finds $\sigma_{\pm} = \frac{d - a \pm \sqrt{\left( a-d\right) ^2 + 4bc}}{2c}$. Moreover, the eigenfunctions will be smooth provided $c(t,x)$ 
does not vanish in some spacetime point. 
Therefore we conclude that the two 
eigenfunctions are $w_\pm= \Pi + \sigma_{\pm} Q$  with the corresponding eigenvalues 
$\lambda_{\pm} = a+\sigma_\pm c=\frac{a + d \pm \sqrt{\left( a-d\right) ^2 + 4bc}}{2}$. 
The eigenvalues are real and the eigenfunctions are 
non degenerate if $\left( a-d\right) ^2 + 4bc >0$ and $c\neq 0$. If this condition 
holds and the eigenfields are smooth then the system is strongly hyperbolic 
\footnote{If the function $c(t,x)$ is globally null,
the equation for $Q$ decouples and $Q$ is itself an eigenfunction propagating with 
speed $d$; the other eigenvalue and eigenfunction can be obtained 
by proceeding in the same way, directly by putting $c=0$ in Eq. (\ref{jg1}). 
It gives $w_2= \Pi + bQ/(a-d)$ (with $a\neq d$) and $\lambda_2= a$. However, if $c=0$, $a=d$ and $b\neq 0$ then 
the system degenerates and becomes weakly hyperbolic. In such a case 
the characteristic matrix is defective like a {\it Jordan block} (cf. Eq. \ref{toysys}) which 
implies that it cannot be diagonalized. A similar situation happens if $b=0$, $a=d$ and $c\neq 0$. On the other hand, 
if $c=0=b$ then both equations (\ref{toy1}) and (\ref{toy2}) decouple and therefore $\Pi$ and $Q$ are 
themselves eigenfunctions (or any linear combination of them is an eigenfunction as well if in addition $a=d$).}. 

Now, let us find the eigenfunctions of the same system using the standard method of linear algebra. We rewrite the system in a more convenient matrix form as:
\begin{equation}
\label{toysys}
\partial_t \left( \begin{array}{cc} \Pi \\ Q \end{array}\right) + \left( \begin{array}{cc} a &b \\ c &d \end{array}\right) \partial_x \left( \begin{array}{c} \Pi \\ Q \end{array}\right) \simeq 0 \,\,\,.
\end{equation}
The eigenvalues $\lambda$ of the characteristic matrix turn to be 
$\lambda_{\pm} = \frac{a + d \pm \sqrt{\left( a-d\right) ^2 + 4bc}}{2}$, which are identical to the ones found 
before. On the other hand, the eigenvectors are:
\begin{equation}
v_{\pm}= \left( \begin{array}{c} \frac{d - a \mp \sqrt{\left( a-d\right) ^2 + 4bc}}{2c} \\ -1 \end{array}\right) 
\,\,\,.
\end{equation}
Note that the first component is identical to $\sigma_\mp$. Thus, the diagonalizing matrix reads 
\begin{equation}
 \mathbb{R} = \left( \begin{array}{cc} \sigma_{-} &\sigma_{+}\\ -1 &-1 \end{array}\right)\,\,\,.
\end{equation}
Then we can write the eigenfunctions as:
\begin{equation}
\left( \begin{array}{c} f_1\\ f_2 \end{array}\right) = \mathbb{R}^{-1}\left( \begin{array}{c} \Pi\\ Q \end{array}\right) \ = \frac{1}{\sigma_{+} - \sigma_{-}} \left( \begin{array}{c} -\Pi - \sigma_{+} Q \\ \Pi + \sigma_{-} Q \end{array}\right) 
\end{equation}
which are the same found before using the judicious guessing approach modulo factors  
$\frac{\mp 1}{\sigma_{+} - \sigma_{-}}=\mp c/\sqrt{\left( a-d\right) ^2 + 4bc} $. The fact that we 
do not obtain exactly the same eigenfunctions is because the eigenvectors are unique up to 
a rescaling. However, both couples of eigenfunctions $w_\pm$ and $f_{1,2}$ are physically 
equivalent. Finally, notice that if $a=d$ then 
$\lambda_{\pm} = a \pm \sqrt{bc}$, and $w_\pm= \Pi \pm \sqrt{b/c}\,Q$
thus `$a$' will play the same roll as 
$-\beta^i$ in the field equations of Sec. 2. For instance, if 
the coefficients $b,c$ are null, the equations (\ref{toy1}) and (\ref{toy2}) decouple 
and then $\Pi$ and $Q$ are themselves eigenfunctions, indicating that they propagate 
along the ``time lines'' (with speed $a$). On the other hand, when $b$ and $c$ are 
non null, these coefficients are related to the 
propagation of the eigenfields 
along the ``light cones''. For instance, $b$ and $c$ can be identified with 
 $\alpha \gamma^{xx}$ and 
$\alpha$, respectively, in many of the field equations (assuming propagation only in the `$x$' direction). 
Thus $\lambda_{\pm}$ are identified with the speeds of propagation $-\beta^x \pm \alpha \sqrt{\gamma^{xx}}$ 
(along the line cones) of many of the eigenfields  
used in the hyperbolicity analysis of STT [cf. Eq. (\ref{Hyp}) ] \cite{Martinez07} or even GR \cite{Alcubierre08} 
(e.g. $w_\pm= u_1 \pm \sqrt{\gamma^{xx}} u_2$ 
where $u_1$ and $u_2$ represent two field variables of STT [cf. Eq. (\ref{PDE}) ]).


\end{document}